\documentclass[12pt]{amsart}
\newtheorem{theorem}{Theorem}[section]
\newtheorem{lemma}[theorem]{Lemma}

\theoremstyle{definition}

\theoremstyle{remark}

\numberwithin{equation}{section}

\def\H{\mathcal H}
\def\Ko{\mathcal K}
\def\RE{\mathbb R}
\def\CO{{\mathbb C}}
\def\HB{\bar{\mathcal H}}
\def\B{\bar B}
\def\K{\mathsf K}
\def\U{\mathsf U}
\def\W{\mathsf W}
\def\uno{\mathsf 1}
\def\Ham{\mathsf H}
\def\C{\mathsf C}
\def\S{\mathsf S}
\def\e{\mathsf e}
\def\G{\mathcal G}

\def\dphi{\dot\phi}
\def\Ei{\text{\rm Ei}}

\begin{document}

\title[Dynamics and Radiation for a Model in QED] {Rigorous Dynamics
  and Radiation Theory for a Pauli-Fierz Model in the Ultraviolet
  Limit}

\author{Massimo Bertini}

\address{Dipartimento di Matematica, Universit\`a di Milano, I-20133
Milano, Italy}

\email{bertini@mat.unimi.it}

\author{Diego Noja}

\address{Dipartimento di Matematica e Applicazioni, Universit\`a Di
Milano-Bicocca, I-20126, Milano, Italy}

\email{noja@matapp.unimib.it}
\author{Andrea Posilicano}

\address{Dipartimento di Scienze Fisiche e Matematiche, Universit\`a dell'Insubria, I-22100
Como, Italy}

\email{posilicano@uninsubria.it}

\begin{abstract}The present paper is devoted to the detailed study of 
  quantization and evolution of the point limit of the Pauli-Fierz
  model for a charged oscillator interacting with the electromagnetic
  field in dipole approximation. In particular, a well defined
  dynamics is constructed for the classical model, which is
  subsequently quantized according to the Segal scheme. To this end,
  the classical model in the point limit, already obtained in \cite{[NP2]}, is
  reformulated as a second order abstract wave equation, and a
  consistent quantum evolution is given. This allows a study of
  the behaviour of the survival and transition amplitudes for
  the process of decay of the excited states of the charged particle,
  and the emission of photons in the decay process. In particular,
  for the survival amplitude the exact time behaviour is found. This
  is completely determined by the resonances of the systems plus a tail
  term prevailing in the asymptotic, long time regime. Moreover, the
  survival amplitude exhibites in a fairly clear way
  the Lamb shift correction to the unperturbed frequencies of the
  oscillator. 
\end{abstract}

\maketitle

\section{Introduction}
In recent years a considerable effort was tributed by the mathematical
physics community to the problem of a rigorous formulation of the
dynamics of the main models in nonrelativistic quantum field theory.
In particular, a comprehensive study of the Pauli-Fierz model, the
model which describes the low energy interaction of nonrelativistic
matter and electromagnetic radiation, was undertaken
by various authors, both in its full form, or making resort to different approximations, such as rotating wave approximation, dipole approximation, or others.
Correspondingly, a wealth of results concerning various aspects
of the model have been obtained, concerning self-adjointness
of the hamiltonian, existence, multiplicity or also non existence of
the ground state and related infrared behaviour, and detailed study of
the spectral properties of the model and of its resonances.
In the present paper we
  give a comprehensive analysis of some of the previous problems in the
  special case of a {\it point} charged oscillator interacting with
  the electromagnetic field in dipole approximation. While this model
  is unrealistically simple compared to the case, to give an example,
  of the hydrogen atom interacting with the full (not dipole)
  radiation field, (about which a lot is known thanks to the work of
  Lieb et al. \cite{[GLL]},\cite{[LL]} and Fr\"olich et
al. \cite{[BFS1]},\cite{[BFS2]}, the novelty of the present work resides in
  the fact that we are able to cope with the point limit of the model.
  The removal of the ultraviolet cutoff in the interaction between
  matter and radiation is in its generality, a difficult and unsolved
  problem, and in particular one not faced off in the quoted rigorous
  literature. In some previous papers of the last two authors
  (\cite{[NP1], [NP2]} and references therein) the {\it renormalized}
  dynamics of the classical Pauli-Fierz model in dipole approximation
  (\cite{[Paulifierz]}) and for fairly general external potentials, 
 was rigorously constructed and analyzed. In particular, it
  was shown that the evolution of the Pauli-Fierz model in the point
  limit is given by an abstract wave equation generated by a family of
  operators related to the so called point interactions
 (see \cite{[AGHH]} and references therein). 
Given the classical model in the ultraviolet limit,
  a second step would be to construct the quantized model. The
  harmonic potential has the unique feature of giving rise to linear
  equations of motion, so that the classical equations for the system
  take the form of an abstract {\it linear} wave equation. This allows a
  plain quantization {\it \`a la} Segal of the model, which seems
  otherwise quite problematic (in contrast with the case
  of the regularized Pauli-Fierz model, where canonical quantization
  works: a detailed study of the regularized dipole Pauli-Fierz model with an
  external harmonic potential was given by Arai in \cite{[Arai]}). So
  we confine ourself to this case, about which we give in the next
  section a self-contained treatment independent and different in spirit 
from \cite{[NP2]}, to clarify some of the themes discussed above.
We outline briefly our
  main results. 
The classical evolution of the system is given by the
  abstract wave equation 
$$
({\partial^2_{tt}}+L_\e) \xi=0
$$
for the couple $\xi=(A,p)$ where
  $A$ is the vector potential of the electromagnetic field in Coulomb
  gauge and $p$ is the particle momentum variable (see section 2 for the
  explanation of this choice). 
Here $L_\e$ is a self-adjoint operator in $L^2_*(\RE^3)\oplus\RE^3$ ($_*$
  stands for ``divergenceless'') such that its resolvent can be 
explicitly calculated (see Lemma 2.2 and Theorem 2.3).
\par 
The operator $L_\e$ has an single
negative eigenvalue and the rest of its spectrum is purely absolutely
continuous and coincides with $[0,+\infty)$. Thus, to quantize the abstract
wave 
equation above 
according to the Segal method, one has to take $L_\e^+$, i.e. $L_\e$ 
projected onto the spectral subspace corresponding to
$\sigma_{ac}(L_\e)$. The (first) quantized
dynamics of the system is defined through the Schr\"odinger-like equation
$i\dot\psi=(L_\e^+)^{1/2}\psi$, defined on the complex Hilbert
space $L^2_*(\RE^3;\CO^3)\oplus\CO^3$ and the hamiltonian of the
quantized system of particle and field is given by the second
quantization $d\Gamma ((L_\e^+)^{1\over 2})$, on the Fock space over
$L^2_*(\RE^3;\CO^3)\oplus\CO^3$.
\par A preliminary but essential step
in the description of the properties of the system is to write the
first quantized evolution in terms of the resolvent of the original
{\it classical } operator $L_\e^+$. 
By Stone formula, spectral theorem and after some work, one
gets (see Lemma 2.5) a representation (perhaps new or at least 
not known to us) for the transition amplitudes 
$\langle\psi_1,e^{-it(L^+_\e)^{1/2}}\psi_2\rangle$ 
between one particle states in terms of the resolvent of $L_\e$. 
Correspondingly, one has an
expression for the survival and transition amplitudes for the second
quantized model on the Fock space, just using functoriality of
$\Gamma$. Our main concern here is in the calculation of two relevant
characteristics of the evolution. The first is the survival amplitude 
$\mathcal S(t)$ of the unperturbed first excited bound state of the 
oscillator. The
second is the amplitude transition $\mathcal A(t)$ between states
with the photon field in the vacuum state and the oscillator in the first
excited state, and states with one photon and the
oscillator in the ground state. The survival and transition amplitudes
relative to more general states factorize in a sum of product of these
two simpler types.  The
survival amplitude $\mathcal S(t)$ 
has a particularly neat form (see Theorem 3.3):
$$
\mathcal S (t)=c_1 e^{-\lambda_\e |t|} +c_2 e^{-\gamma_\e |t|}
e^{-i\omega_\e t} + c_3 e^{-\lambda_\e |t|} \Ei(\lambda_\e |t|)\,.
$$
In this formula, the complex numbers
$$
z_k:=(-1)^{k}\omega_\e-
i\gamma_\e\,,\quad k=1,2
$$
coincide with the complex poles of the analytically continued
resolvent of $L_\e^+$; $-\lambda_\e^2$ is the unique negative eigenvalue
of $L_\e$, and by $\Ei$ we mean the exponential integral function.
The complex numbers $c_1$, $c_2$ and $c_3$ depend on the physical
parameters.
So, in the time evolution of the survival amplitude for
the bound states of the oscillator it is possible to distinguish three
different time behaviours. The first term, depends on a resonance on the 
imaginary axis originated from the projection onto the positive spectral subspace; 
it is a pure exponentially decaying term (for
positive times) and the characteristic time of the decay, for
realistic values of mass and charge of the electron, has an order of $
10^{-23}$ sec, an exceedingly small time.  The second term is an
exponentially damped oscillation described in terms of the complex
resonance poles. In particular, the imaginary part of the resonance
$\gamma_\e$ gives as usual a measure of the lifetime of the excited
unperturbed states, or, equivalently, the breadth of the spectral
lines of the spontaneous decay of the excited states (according to 
the Breit Wigner law, see \cite{[BreitWigner]}); while the real
part gives the position of the maximum in the emission of the spectral
line. In terms of given physical parameters of the systems, the
behaviour of these quantities is the following:
$$
\omega_\e=\omega_0+\frac{28\omega_0^3}{3m^2c^6}\,\e^4+O(\e^5)\,,\qquad
\gamma_\e=\frac{2\omega_0^2}{3mc^3}\,\e^2+O(\e^5)
$$
Here the symbols $m$, $c$ and
$\e$ denote the (renormalized) 
phenomenological mass, the velocity of light and the electric charge
respectively. The noteworthy fact is the Lamb shift in the expression of
$\omega_\e$. The maximum in the emission does not appear in
correspondence of the unperturbed frequency of the oscillator, but at
a displaced frequency.  The last term, taking into account the asymptotic behaviour of the exponential integral, is of the
order $\frac{1}{t}$ for $t\gg 1$.  The appearance of a slowly decaying 
tail implies a departure from the purely exponential decay given by the Breit 
Wigner law, and is well known both in theoretical models and experimental studies.  
See Lemma 2.5 and the following remarks for an interpretation of its origin.\par\noindent
The transition amplitudes $\mathcal A(t)$ have a form
very similar to the one for $\mathcal S(t)$ (see Theorem 3.5), and to them 
apply the same remarks and comments concerning their time behaviour.

\section{An Exactly Soluble Model in Classical and Quantum Electrodynamics}
\subsection{Classical theory}
The classical Pauli-Fierz model for a particle with charge $\e$,
charge density $\e \rho_r$ and bare mass $m_r$ interacting with the
electromagnetic field in dipole approximation and subjected to a
restoring harmonic force, is described by the hamiltonian
$$
H=2\pi c^2  \langle E, E \rangle 
+\frac1{8\pi}\, \langle A,\Delta A\rangle 
+\frac1{2m_r}\,\left| p-\frac {\e}{c}\, \langle\rho_r,  A\rangle 
\right|^2+\frac12\,\alpha  |q|^2\ ,
$$
where $E={\dot A}/(4\pi c^2)$ is the canonical variable conjugated to
$A$ and $p$ is the canonical
momentum conjugated to the particle position $q$. The model is written
in the Coulomb gauge, so the field are divergenceless.  A suitable
phase space for this dynamical system is $H^1_*(\RE^3)\oplus
L^2_*(\RE^3)\oplus \RE^3\oplus \RE^3$ where $H^1_*(\RE^3)$ denotes the
space of divergenceless locally square integrable vector fields with
square integrable first derivatives. 
A point electron should have a
Dirac measure as charge distribution, and the regular form factor
$\rho_r$ is introduced to give meaning to the the equations of motion.
So $r$ has to be interpreted as a measure of the particle radius.
In the point limit, as $r \downarrow 0$, the charge distribution $\rho_r$
weakly converges to $\delta_0$, and the Hamilton equations
corresponding to the Pauli-Fierz hamiltonian loose their
original meaning. A well defined dynamical system is recovered only at
the expence of renormalizing the bare mass $m_r$. 
This procedure is analyzed in detail in \cite{[NP1]} and we content ourself to
say here that the correct prescription is given by the
well known relation between the bare, electromagnetic
($m_{em}$) and
renormalized ($m$) masses,
$$
m=m_r + m_{em}=m_r + {{8\pi}\over 3}{{\e^2}\over {c^2}}\,\langle
(-\Delta)^{-1} \rho_r,\rho_r \rangle \ .
$$
Keeping $m$ fixed to the physical value, this is the only choice
for the bare mass which allows to obtain a nontrivial limit for the
Pauli-Fierz model. From now on, with the symbol $m_r$ we mean precisely the function
$m_r(\e, m)$ given by the above relation.
\par
The rigorous deduction of the limit dynamics is carried out in \cite{[NP2]}
for a general external potential; specializing the result to the case
of the harmonic potential one obtains the equations
\begin{align*}
\dot A&=4\pi c^2E\cr
\dot E&=-\frac{1}{4\pi}\,H_m^pA\cr
\dot q&=Q_A\cr
\dot p&=-\alpha q\,.
\end{align*}
The $p$-dependent operator $H_m^p$ is an affine deformation of a linear 
self-adjoint operator in the class of point
interactions (see \cite{[AGHH]} and references therein for the use of
point interactions in quantum mechanics): $H^p_m$ is
linear if and only if $p=0$ and $H^0_m$ is the vector-valued
version of one of such linear operators; $Q_A$ is a certain linear 
functional which in some sense extract the singular part of the vector potential.
 \par
While obtaining a well defined dynamics in the point limit it is an 
interesting 
and not obvious result, it remains unclear how to quantize such a system.
So, according to the point of view we adopt in this paper, and in view of the
Segal quantization of the system, we would like to work with (abstract)
second order wave equations of the form
$$
\ddot\xi(t)=-L \xi(t)\ ,
$$
with a suitable self-adjoint operator $L$. This is not the case of the
previous system of equation, due to the presence of a tight relation
between the dynamical variables contained in the definition of the
operator $H_m^p$. Going back to the regularized system, the
corresponding second order equations are not better from this point of
view; they are 
\begin{align*}
&\frac1{c^2}\ddot  A= \Delta A+ 
\frac{4\pi \e} c\, M\dot q \rho \ ,
\cr
&m_r\ddot q =-{\frac{\e}c }\, \langle\rho_r,\dot A\rangle
-\alpha q\,, 
\end{align*}
and the presence of $\dot q$ and $\dot A$ on the right hand side make
them not of the desired form.
So we prefer to give an construction of the limit
dynamics independent of the one given in \cite{[NP2]}.  In fact, making resort
to the hamiltonian regularized equations above,
it is simple to overcome this problem. Deriving with respect to time
the last equation and using the third, one obtains that the couple
$\xi=(A,p)$ satisfies the abstract second order wave equation
$$
\ddot\xi=-L^r_{\e} \xi
$$
where the operator $L^r_{\e}$ is given by
$$
L^r_{\e}(A,p)
=\left(-c^2\Delta  A-\frac{4\pi \e c}{m_r}\,M\left(p-\frac 
{\e}{c}\,\langle \rho_r, A\rangle\right)\rho_r , 
\frac{\alpha}{m_r}\,\left(p-\frac 
{\e}{c}\langle \rho_r, A\rangle
\right)\right)$$
The analogous result is obtained by means of the canonical
transformation which exchanges the particle position and momentum (see
also \cite{[Spohn]})
$$
q=- P\ ,\quad  p= Q \,.
$$
The change of dynamical variable from $q$ to $p$ in the second order (or
lagrangian) formalism is not particularly relevant to
the analysis of the problems we are interested in, and the
interpretation of the results we get. For example, the time behaviour
of the classical position $q$, which is important in the calculation
of the survival amplitude of the bound states of the oscillator, is
quite simply related to the time behaviour of the variable $p$: up to
a constant, the derivative of the momentum gives the position, and
this relation is preserved in the limit dynamics.\par
Another important point to note, is that the operator $L^r_{\e}$ is a finite
rank perturbation of the noninteracting operator $L_0=(-c^2\Delta, 0)$.
This simple structure suggests the possibility that the operator
$L^r_{\e}$ has a limit for $r\downarrow 0$, and being an unbounded operator
such a limit, if existing, should be sought in the resolvent sense.
The calculation of the resolvent of the operator $L^r_{\e}$ is lenghty
but elementary. We omit the proof and give the result in the following 
\begin{lemma} For every $z\in\CO_\pm$, the resolvent of 
$L^r_{\e}$ is given by 
$$
(L^r_\e-z^2)^{-1}(A,p)=(\G_z^\pm *A,0)-\Lambda^r(z)\,R^r_\e(z)(A,p)\,,
$$
where
$$
\Lambda^r(z)= \frac{-1}{m_rk_1^rk^r_2}\, ,\quad 
\G_z^\pm(x)=\frac{1}{c^2}\,\frac{e^{\pm
i{z}\,|x|/c}}{4\pi|x|}\,,\quad\text{\rm $\pm$Im$z>0$}\,, 
$$
$$
R^{r}_\e(z)(A,p)= \left(R^{1r}_\e(z)(A,p),R^{2r}_\e(z)(A,p)\right)  \,,
$$
\begin{align*}
R^{1r}_\e(z)(A,p)&=\frac{4\pi \e}{c}\,
M\left(\left({z^2}\,c\,{\e}\,\langle(-c^2\Delta-z^2)^{-1}A,\rho_r\rangle
\right.\right.\\&\quad\left.\left.
+c^2\,{{m_rk^r_1}\over
{\alpha}}\,(z^2+k^r_2) p\right)(-c^2\Delta-z^2)^{-1}\rho_r\right) \\
R^{2r}_\e(z)(A,p)&=\alpha\,\frac{\e}{c}\,\langle(-c^2\Delta-z^2)^{-1}A,\rho_r\rangle+
m_rk^r_1\,p \,,
\end{align*}
\begin{align*}
k^r_1&=1+\frac{8}{3}\pi{\e}^2 \langle(-c^2\Delta-z^2)^{-1}\rho_r, \rho_r\rangle \ ,\\
k^r_2&=\frac{\alpha}{m_r} - z^2- \frac{8}{3}\frac{\alpha\pi{\e}^2}{m_r^2k^r_1}
 \langle(-c^2\Delta-z^2)^{-1}\rho_r, \rho_r\rangle\,.
\end{align*}
\end{lemma}
The next step is the point limit $r\downarrow 0$. The result is
analogous to the corresponding result given in \cite{[NP1]} for the case of a
free particle, and the proof is modelled on one of the well known ways
of defining point interactions (see \cite{[AGHH]}). 
We give only an outline of the proof.
\begin{lemma} Let $\rho_r\to\delta_0$ weakly as $r\downarrow 0$.
For every fixed $z\in\CO_\pm$, $(L^r_\e-z^2)^{-1}$ converges 
as $r\downarrow 0$ in the norm resolvent sense to the operator
$$(\G_z^\pm *A,0)-\Lambda_\pm(z)R^\pm_\e(z)(A,p)\,,$$
where, putting $\omega_0^2:=\frac{\alpha}{m}$,
\begin{align*}
R^\pm_\e(z)(A,p)=&\left(
\frac{4\pi \e}{c}\,
M\left({z^2}\,c\,{\e}\,\langle\G^\mp_{z^*},A\rangle+c^2p\right)
\G_z^\pm,\right.\\
&\ \left.m\omega_0^2\,
\frac{\e}{c}\,\langle\G_{z^*}^\mp,A\rangle+
\left(m\pm i\,\frac{2\e^2}{3c^3}\,z\right)\,p\right)\,,
\end{align*}
$$
\Lambda_\pm(z)=\frac{1}{
\pm i\,\frac{2\e^2}{3c^3}\, z^3-m(\omega_0^2-z^2)}\,.
$$
Such an operator is the resolvent of a self-adjoint operator $L_\e$ on
$L^2_*(\RE^3)\oplus\RE^3$ when on the component $\RE^3$ one considers the scalar product
$\langle p_1,p_2\rangle:=\kappa_0\,p_1\cdot p_2$, with
$\kappa_0:=\frac{4\pi c^2}{m\omega_0^2}$.
\end{lemma}
\begin{proof}
  The proof of the convergence of the regularized resolvent is a
  direct consequence of the following limiting relations:
$$\lim_{r\downarrow 0}\,(-c^2\Delta-z^2)^{-1}\rho_r=\G_z^\pm\,,$$
$$
\lim_{r\downarrow 0}\,k^r_1=0\,, \qquad 
\lim_{r\downarrow 0}\,k^r_1 m_r=m\pm i\,\frac{2\e^2}{3c^3}\,z\,,
$$
$$
\lim_{r\downarrow 0}\,k^r_2=-{{\pm i\,\frac{2\e^2}{3c^3}}\,
  {z^3-m(\omega_0^2-z^2)}\over {m\pm i\,\frac{2\e^2}{3c^3}\,z}}\,.
$$
Checking that the limit operator is the resolvent of
a self-adjoint operator is routine. See \cite{[NP1]} for a similar 
verification. 
\end{proof}
By the resolvent just constructed it is straightforward to derive
the actions of $L_\e$ itself and of its spectral properties:
\begin{theorem}
The action and domain of the self-adjoint operator
$$
L_\e\,:\,D(L_\e)\subseteq L^2_*(\RE^3)\oplus\RE^3\to L^2_*(\RE^3)\oplus\RE^3\,
$$
are given by
$$
L_\e(A,p)=\left(-c^2\Delta
A_0,\,\omega_0^2\left(p-\frac{\e}{c}A_0(0)\right)\right)\,,
$$
\begin{align*}
&D(L_\e)=\left\{(A,p)\in L^2_*(\RE^3)\oplus\RE^3
\,:\, A=A_0+\frac{4\pi\e}{c}\,Mv\,\G\,,\right.\\
&\left. \sqrt{-\Delta}A_0,\ \Delta A_0\in L^2_*(\RE^3)\,,
\ v\in\RE^3\,,\ mv=p-\frac{\e}{c}A_0(0)\right\}\,,
\end{align*}
where $$\G(x)=\frac{1}{4\pi|x|}\,.$$
Moreover,
$$
\sigma_p(L_\e)=\left\{-\lambda_\e^2\right\}\,,\quad\sigma_{ess}(L_\e)
=\sigma_{ac}(L_\e)=[0,+\infty)\,,\quad\sigma_{sc}(L_\e)=\emptyset\,,
$$
where $$\lambda_\e=\frac{3mc^3}{2\e^2}+O(\e^2)$$ 
is the unique real (and positive) solution of the third order
equation
$$
\frac{2\e^2}{3c^3}\,\lambda^3-m(\omega_0^2+\lambda^2)=0
\,.
$$
\end{theorem}
We emphasize that, due to the fact that the operator $L_\e$ has been
constructed as a norm resolvent limit of the operator $L^r_\e$, the
flow generated by the limit operator coincides with the limit of the
regularized flow, in the relevant norms of the phase space and
uniformly in time.  This allows to consider $L_\e$ as the generator of
the limit dynamics.  As a second remark, note that the algebraic
equation $\Lambda_\pm(z)^{-1}=0$ is nothing but the characteristic
equation of the Abraham-Lorentz equation (see e.g. \cite{[Lorentz]}) 
in the particular case of an harmonic external force, i.e.
$$
-\tau_0 \dddot q + \ddot q + \omega_0 q=0  \qquad\qquad 
\tau_0={{2\e^2}\over{3mc^3}}\ ,
$$
the equation which classically describes the behaviour of the particle
position in the nonrelativistic regime (the classical relativistic equation was
obtained by P.A.M.Dirac in \cite{[Dirac]}). In particular, the
negative eigenvalue of $L_\e$ corresponds to the so called runaway
solution of the A-L equation. In the traditional approaches, this
solution is discarded due to its unphysical character. From the
present point of wiew, we give up the (important) interpretative
problem related to the presence of these instabilities, and take the
attitude according to which the suppression of runaway behaviour
corresponds to reduction of the dynamics on the stable subspace, or
equivalently, restriction to the absolutely continuous component of
the spectrum.  This should correspond, in ordinary scattering theory
for Schr\"odinger operators, to the elimination of bound states. The
procedure to obtain this reduction can be explicitely performed as
follows. Let us consider, from now on, $L_\e$ as acting on the complex Hilbert
space $L^2_*(\RE^3;\CO^3)\oplus\CO^3$. It is self-adjoint when on the 
component $\CO^3$ one considers the scalar product
$\langle\zeta_1,\zeta_2\rangle:=\kappa_0\,\zeta_1^*\cdot \zeta_2$.
\par
Given $\zeta_1,\, \zeta_2,\,\zeta_3$, an orthonormal base in $\CO^3$,
and defining
$$
\G_{\lambda_\e}(x):=\frac{1}{c^2}\,\frac{e^{-\lambda_\e\,|x|/c}}{4\pi|x|}\,,
\quad
\kappa:=\left(\frac{2\pi\e^2\lambda^3_0}{m^2\omega_0^4c}+\kappa_0\right)^{1/2}
\,,
\quad \kappa_1:=\,\frac{\e}{c}\,\lambda_\e^2\kappa_0\,,
$$
let 
$$\psi_i^0=\frac{1}{\kappa}(-\kappa_1
M\zeta_i\G_{\lambda_\e},\zeta_i)\,,\quad i=1,2,3$$
be the normalized eigenvectors corresponding to the eigenvalue
$-\lambda_\e^2$. Then the projection $P_\e$ onto the
absolutely continuous subspace of the operator $L_\e$ is given by 
$$P_\e\psi\equiv\psi^+=
\psi-\sum_{i=1,2,3}\langle\psi,\psi^0_i\rangle\psi^0_i\,.$$
In particular 
$$
P_\e(0,\zeta)=\frac{\kappa_2}{\kappa^2}\,\,
\left(\frac{12\pi c^4}{\e}\,M\zeta
\G_{\lambda_\e},\lambda_\e\zeta\right)
\,,\quad\kappa_2:=\frac{4\pi\e^2\lambda_\e^2}{m^2\omega{_0^4}c}
\,.
$$
We define the positive self-adjoint operator $L_\e^+$ by
$L_\e^+:=P_\e L_\e\ ,$ and from now on we consider this reduced operator 
as the generator of the physical limit dynamics.
\par \noindent 
\subsection{Quantum theory}
The operator $L_\e^+$ generates the
classical evolution, whereas, according to the results summarized in
the appendix, the corresponding quantum evolution is given
in terms of its square root $(L_\e^+)^{1/2}$.  More explicitely,
denoting by ${\mathcal F}(L_*^2(\RE^3;\CO^3)\oplus \CO^3)$
the bosonic Fock space over $L_*^2(\RE^3;\CO^3)\oplus \CO^3$, i.e. 
$$
{\mathcal F}(L_*^2(\RE^3;\CO^3)\oplus \CO^3)
:=\bigoplus_{n\ge 0}S_n (L_*^2(\RE^3;\CO^3)\oplus\CO^3)^{\otimes n}\,,
$$ 
where $S_n$ denotes the symmetrization operator on the $n$-th sector,
the quantum hamiltonian on ${\mathcal F}(L_*^2(\RE^3;\CO^3)\oplus \CO^3)$
corresponding 
to the second quantization of the classical wave equation
$$
\left(\partial^2_{tt}+L_\e^+\right)(A,p)=0
$$ 
is given by $$\Ham_\e:=\hbar\, d\Gamma((L_\e^+)^{1/2})\,.$$
Note that the noninteracting hamiltonian
$$\Ham_0:=\hbar\, d\Gamma(L_0^{1/2})\equiv 
\hbar\, d\Gamma(\sqrt{-\Delta}\oplus \omega_0)$$
is unitarily equivalent to $$\hbar\, d\Gamma(\sqrt{-\Delta})\otimes
\uno+\uno\otimes \hbar\, d\Gamma(\omega_0)\,,$$ 
defined on the Hilbert space
$$
{\mathcal F}(L_*^2(\RE^3;\CO^3))\otimes {\mathcal
F}(\CO^3):=\bigoplus_{n\ge 0}S_n L_*^2(\RE^3;\CO^3)^{\otimes n}\otimes
\bigoplus_{n\ge 0}S_n (\CO^3)^{\otimes n}\,.
$$ 
The unitary operator 
$$
\U:{\mathcal F}(L_*^2(\RE^3;\CO^3)\oplus\CO^3)\to
{\mathcal F}(L_*^2(\RE^3;\CO^3))\otimes {\mathcal F}(\CO^3)\eqno (1)
$$ giving the stated equivalence is defined 
by 
$$
\U\Omega:=\Omega\otimes\Omega\,,\quad \U\C(\psi)\U^{-1}=
\C(\varphi)\otimes \uno+\uno\otimes\C(\zeta)\,, 
$$
where $\Omega$ denotes the vacuum, $\psi=(\varphi,\zeta)$ and $\C$ is
the usual creation operator.
Moreover $\hbar\, d\Gamma(\omega_0)$ is unitarily equivalent to the usual
harmonic oscillator hamiltonian on $L^2(\RE^3;\CO)$ 
given by the self-adjoint operator 
$$
-\frac{\hbar^2}{2m}\,\Delta+\frac{m\omega_0^2}{2}\,q^2\,.
$$ 
Concerning the interacting Hamiltonian $\Ham_\e$, the present one
is its first {\it explicit} construction in the ultraviolet limit and
the problem arises if other representations more directly confrontable
with usual canonical formulation could be given. We emphasize the fact
that in our description, already at the classical level, the point
limit produces an intimate interlacing between field singularities and
particle variables, through the definition of the domain of the
operator $L_\e$ itself, and this fact introduces essential
difficulties in tracing the relation with the canonical formalism
based on the usual regularized Pauli-Fierz Hamiltonian, where this constraint 
disappears. Nevertheless,
in the quoted Arai paper (\cite{[Arai]}), a reconstruction theorem
based on the limit of the Wightman functions of the regularized model
is outlined. It could be interesting to indagate the relations between
two approaches. \par\noindent
Since the resolvent of $L_\e$ is quite explicit,
making use of the Birman-Kato invariance principle to deal with the
group generated by the square root, and of the Birman-Kuroda
completeness theorem (see \cite{[Birman]}) which is applicable because
$(L_\e+z)^{-1}-(L_0+z)^{-1}$ is a finite rank operator, one
immediately obtains the following
\begin{theorem}  
Let $P_{0}$ and $P_\e$ be the orthogonal projections onto the
absolutely continuous subspaces of $L_0$ and $L_\e$ respectively. Then
the M\"oller wave operators exist, they are complete and,
$$
\Omega_\pm(\Ham_\e,\Ham_0)
:=\text{\rm s-}\lim_{t\to\pm\infty}\,
e^{-i{t}\Ham_\e/{\hbar}}\,
e^{-i{t}\Ham_0/{\hbar}}\,\Gamma(P_0)=\Gamma(\Omega_\pm(L_\e,L_0)) \,
$$
$$
\Omega_\pm(\Ham_0,\Ham_\e)
:=\text{\rm s-}\lim_{t\to\pm\infty}\,
e^{-i{t}\Ham_0/{\hbar}}\,
e^{-i{t}\Ham_\e/{\hbar}}\,\Gamma(P_\e)=\Gamma(\Omega_\pm(L_0,L_\e)) 
$$

\end{theorem}
\par \noindent
A final result of this paragraph is a formula for the evaluation of
transition amplitudes of the Schr\"odinger-like propagator
$e^{-it(L^+_\e)^{1/2}}$, that is the scalar product of the type
$\langle\psi_1, e^{-it(L^+_\e)^{1/2}}\psi_2\rangle$,  in terms of
boundary values of the resolvent of the classical operator $L_\e$.
In the formula the special form of the operator $L_\e$ plays
no rule, and it holds true for a positive generator $A$ whatsoever.  By
linearity it will be sufficient to suppose that $\psi_1$ and $\psi_2$
are real valued.  Since (see the appendix for the definition of $W_{L_\e^+}$)
$$
((iW_{L_\e^+}-z)^{-1})(\psi,0)=
(z(L^+_\e-z^2)^{-1}\psi,-i\psi-iz^2(L^+_\e-z^2)^{-1}\psi)\,,
$$
by Lemma A.2 and Lemma A.3 one obtains 
\begin{align*}
&\langle\psi_1,e^{-it(L^+_\e)^{1/2}}\psi_2\rangle\\
=&\lim_{a\uparrow\infty}\lim_{\epsilon\downarrow 0}\,\frac{1}{2\pi i}
\int_{-a}^a d\lambda\ e^{-it\lambda}
(\langle\psi_1,((\lambda+i\epsilon)(L_\e-(\lambda+i\epsilon)^2)^{-1}\\
&\qquad\qquad\qquad\qquad 
-(\lambda-i\epsilon)(L_\e-(\lambda-i\epsilon)^2)^{-1})\psi_2^+\rangle)\\
+&\lim_{a\uparrow\infty}\lim_{\epsilon\downarrow 0}\,\frac{1}{2\pi i}
\int_{-a}^a d\lambda\ e^{-it\lambda}
(\langle\psi_1,((\lambda+i\epsilon)^2(L_\e-(\lambda+i\epsilon)^2)^{-1}\\
&\qquad\qquad\qquad\qquad 
-(\lambda-i\epsilon)^2(L_\e-(\lambda-i\epsilon)^2)^{-1})
(L_\e^+)^{-1/2}\psi_2^+\rangle)
\,.
\end{align*}
Moreover, using first resolvent identity and $$
(L_\e^+)^{-1/2}=\frac{1}{\pi}\int_{\RE} ds\,(L^+_\e+s^2)^{-1}
$$
it turns that
\begin{align*}
&\langle\psi_1,(L_\e-z^2)^{-1}(L^+_\e)^{-1/2}\psi_2^+\rangle\\
=&\frac{1}{\pi}\int_\RE \frac{ds}{s^2+z^2}\,
\langle\psi_1,(L_\e-z^2)^{-1}\psi_2^+\rangle
-\frac{1}{\pi}\int_\RE \frac{ds}{s^2+z^2}\,
\langle\psi_1,(L_\e+s^2)^{-1}\psi_2^+\rangle\\
\end{align*}
Thus for any couple  $\psi_1,\,\psi_2$ for which the limits 
$$
\langle\psi_1,(L_\e-\lambda_\pm^2)^{-1}\psi_2^+\rangle:=
\lim_{\epsilon\downarrow 0}\,
\langle\psi_1,(L_\e-(\lambda\pm i\epsilon)^2)^{-1}\psi_2^+\rangle
$$
exist, one obtains the following
\begin{lemma}
\begin{align*}
&\langle\psi_1,e^{-it(L^+_\e)^{1/2}}\psi_2\rangle\\
=&\lim_{a\uparrow\infty}\,\frac{1}{2\pi i}
\int_{-a}^a d\lambda\ e^{-it\lambda}\,\lambda\,
(\langle\psi_1,((L_\e-\lambda_+^2)^{-1}-
(L_\e-\lambda_-^2)^{-1})\psi_2^+\rangle)\\
+&\frac{1}{\pi}\,\int_\RE 
\frac{ds}{2\pi i}\,
\int_\RE d\lambda\, \frac{e^{-it\lambda}\,\lambda^2}{s^2+\lambda^2}\,
\langle\psi_1,((L_\e-\lambda_+^2)^{-1}-
(L_\e-\lambda_-^2)^{-1})\psi_2^+\rangle\,.
\end{align*}
\end{lemma}
Note that the second contribution in the previous formula comes from
the nonlocal relation between the real phase space classical
variables and the complexified ones, described in Lemma A.3.
Performing however the $s$ integral one obtains, under the same condition of the previous result, the following alternative representation 
\begin{align*}
&\langle\psi_1,e^{-it(L^+_\e)^{1/2}}\psi_2\rangle\\
=&\lim_{a\uparrow\infty}\,\frac{1}{2\pi i}
\int_{0}^a d\lambda\ e^{-it\lambda}\,\lambda\,
(\langle\psi_1,((L_\e-\lambda_+^2)^{-1}-
(L_\e-\lambda_-^2)^{-1})\psi_2^+\rangle)\\
\end{align*}
This last representation has a more direct meaning, in that it is an
integral extended over the spectrum of the operator $(L^+_\e)^{1/2}$.  
Moreover, it presents the
evolution generated by $(L^+_\e)^{1/2}$ as a Fourier transform of a
function (for every fixed couple of states $\psi_1$ and $\psi_2$)
supported on a half line. As a consequence of the Paley-Wiener
theorem, the evolution of the amplitude transition cannot have a
leading large time contribution of exponential type, but a slower one
should appear (see \cite{[Khal1]},\cite{[Khal2]} for an early application of this
remark to the time decay of amplitude transition). The exact time
behaviour cannot be precised without the knowledge of further details
about the generator. To this end it is of more practical use the formula given in Lemma 2.5, as we see in the following section.


\section{Radiation theory}
\subsection{Generalities}
In this section we want to give some details of the quantum dynamics of 
the model we are studying. In particular, we want to estimate, under the
dynamics generated by $\Ham_\e$, the survival amplitudes of the bound
states of $\Ham_0$ and the probability amplitudes of the transitions
between two of such states with emission of photons.\par\noindent
We begin with some preliminaries on the general structure of the 
amplitude transitions in our model.\par
In the space 
${\mathcal F}(L_*^2(\RE^3;\CO^3))\otimes {\mathcal
F}(\CO^3)$ a bound state ($n$-level) for the hamiltonian operator 
$\Ham_0$ is represented
by a vector of the kind $$\Omega\otimes
S_n(\zeta_1\otimes\cdots\otimes\zeta_n)\,.$$ According to (1) this
state is represented in ${\mathcal F}(L_*^2(\RE^3;\CO^3)\oplus\CO^3)$
by the vector $$S_n((0,\zeta_1)\otimes\cdots\otimes(0,\zeta_n))\,$$ 
By (1) again, a more general state with $m$ photons, of the kind
$$S_m(\varphi_1\otimes\cdots\otimes\varphi_m)\otimes
S_n(\zeta_1\otimes\cdots\otimes\zeta_n)\,,$$ 
is represented in ${\mathcal F}(L_*^2(\RE^3;\CO^3)\oplus\CO^3)$ 
by the vector
$$S_{m+n}((\varphi_1,0)\otimes\cdots\otimes(\varphi_m,0)
\otimes (0,\zeta_1)\otimes\cdots\otimes(0,\zeta_n))\,.$$ 
Note that, being $\Ham_\e=\hbar\, d\Gamma((L_\e^+)^{1/2})$, any sector in 
${\mathcal F}(L_*^2(\RE^3;\CO^3)\oplus\CO^3)$ is preserved under
$e^{-it\Ham_\e}$. Thus only the transitions from a $n$-level to a
$m$-level with the emission of $n-m$ photons are allowed.\par 
On the other hand, as we shall see, asymptotically the particle part of the 
wavefunction of the state relaxes to the ground state, and this yelds to the 
asymptotic conservation of the photon number in the scattering process.
This property was devised by A.Arai in \cite{[Arai]} for the
regularized Pauli-Fierz model with quadratic potential, as a consequence of 
the factorization properties of the scattering matrix, and we find it again 
in the point model.\par
Since
\begin{align*} &e^{-it\Ham_\e/\hbar}\,
S_n(\psi_1\otimes\cdots\otimes\psi_n)
  =\Gamma(e^{-it(L^+_\e)^{1/2}})\,S_n(\psi_1\otimes\cdots\otimes\psi_n)\\
  =&\frac{1}{n!}\sum_{\sigma}e^{-it(L_\e^+)^{1/2}}\psi_{\sigma_1}
  \otimes\cdots\otimes e^{-it(L^+_\e)^{1/2}}\psi_{\sigma_n}\,,
\end{align*}
the survival amplitude of the bound state $\Omega\otimes S_n(\zeta_1\otimes\cdots\otimes\zeta_n)$ is given by
$$
\frac{1}{n!}\sum_{\sigma}\prod_{j=1}^n\langle(0,\zeta_{\sigma_j}), 
e^{-it(L^+_\e)^{1/2}}(0,\zeta_j)\rangle
$$
whereas the the probability amplitude of the transition  
$$
\Omega\otimes S_{n}(\zeta_{1}\otimes\cdots\otimes\zeta_{n})
\ \leftrightarrow\ S_m(\varphi_1\otimes\cdots\otimes\varphi_m)\otimes 
S_{n-m}(\xi_{m+1}\otimes\cdots\otimes\xi_{n-m})$$ is given by
$$
\frac{1}{n!}\sum_{\sigma}\prod_{\sigma_j\le m}\langle(\varphi_{\sigma_j},0), 
e^{-it(L^+_\e)^{1/2}}(0,\zeta_j)\rangle
\prod_{\sigma_k> m}
\langle(0,\xi_{\sigma_k}), e^{-it(L^+_\e)^{1/2}}
(0,\zeta_k)\rangle\,.
$$

\subsection{Survival amplitudes}
After these general remarks, we evaluate the survival amplitudes
for the bound states of the unperturbed dynamics along the perturbed
evolution. To simplify the exposition we break the analysis in a number 
of lemmata.
\begin{lemma} 
For any $\zeta_1,\,\zeta_2\in \CO^3$ one has 
$$
\langle(0,\zeta_1),e^{-it(L^+_\e)^{1/2}}(0,\zeta_2)\rangle
=\mathcal S(t)\,\zeta_1^*\cdot\zeta_2 \,,
$$
where $$\mathcal S(t)=-\frac{2\kappa_0\kappa_2}{\kappa^2}\,I(t)$$ and, with $\pm t>0$,
\begin{align*}
I(t)&=\mp e^{\mp
\lambda_\e t}\,i\lambda_\e\,\frac{p(-i\lambda_\e)}{q(-i\lambda_\e)}
\pm e^{\mp \gamma_\e t}\,
\frac{e^{-i\omega_\e t}2z_\pm}{z_\pm+i\lambda_\e}\,
\frac{p(z_\pm)}{q'(z_\pm)}
+J_1(t) + J_2(t)\,,
\end{align*} 
\begin{align*}
&J_1(t) = \frac{1}{\pi}\int_\RE ds\,
\frac{e^{\mp|s|t}|s|}{2(|s|+\lambda_\e)}\,
\frac{p(i|s|)}{q(i|s|)}\,,\\
&J_2(t) = \frac{1}{\pi}\int_\RE ds
\left(\,\frac{e^{\mp|s|t}|s|}{2(|s|-\lambda_\e)}\,
\frac{p(-i|s|)}{q(-i|s|)}
-\frac{e^{\mp\lambda_\e t}\lambda^2_\e}{s^2-\lambda_\e^2}\,
\frac{p(-i\lambda_\e)}{q(-i\lambda_\e)}\,\right)\,,\\
\end{align*}
$$
p(z)=
\frac{\e^2\lambda_\e}{3c^3}\,z+i\,\frac{m}{\lambda_\e}\,
\left({\omega_0^2}+\frac{\lambda_\e^2}{2}\right)\,,\quad
$$
$$
q(z)= 
\frac{2\e^2}{3c^3}\, z^2+i\,\frac{m\omega_0^2}{\lambda_\e^2}\,
(z+i\lambda_\e)\,,
$$
where
$$
z_\pm:=\pm\omega_\e-i\gamma_\e\,,$$
$$
\omega_\e=\omega_0+\frac{28\omega_0^3}{3m^2c^6}\,\e^4+O(\e^5)\,,\qquad
\gamma_\e=\frac{2\omega_0^2}{3mc^3}\,\e^2+O(\e^5)
$$
are the two roots of $q(z)$.
\end{lemma}
\begin{proof} 
  Let $z\in \CO_\pm$. Taking in account the algebraic equation
  satisfied by $\lambda_e$ and the identity 
$$
\langle\G_{z^*}^\mp,\G_{\lambda_\e}\rangle=\frac{1}{4\pi c^3} 
\frac{1}{\lambda_\e \mp iz}
$$ 
we have, for the projected resolvent,
\begin{align*}
&\langle(0,\zeta_1),(L_\e-z^2)^{-1}(0,\zeta_2)^+\rangle\\
=&-\kappa_0\,\frac{\kappa_2}{\kappa^2}\,\Lambda_\pm(z)\,
\left(m\omega_0^2\,8\pi c^3
\langle\G_{z^*}^\mp,\G_{\lambda_\e}\rangle+
\left(m\pm i\,\frac{2\e^2}{3c^3}\,z\right)\,\lambda_\e\right)
\,\zeta_1^*\cdot\zeta_2\\
=&\frac{-2\kappa_0\kappa_2}{\kappa^2(\pm iz-\lambda_\e)}\,\frac{
\left(\pm i\,\frac{\e^2\lambda_\e}{3c^3}\,z-
\frac{m}{\lambda_\e}\,\left({\omega_0^2}+
\frac{\lambda_\e^2}{2}\right)\right)(\pm iz+\lambda_\e)}
{\left(\frac{2\e^2}{3c^3}\, z^2+\frac{m\omega_0^2}{\lambda_\e^2}\,
(\pm iz-\lambda_\e)\right)(\pm iz+\lambda_\e)}
\,\zeta_1^*\cdot\zeta_2\,.
\end{align*}
Thus using the representation formula given in lemma 2.5,  we get
$$
\langle(0,\zeta_1),e^{-it(L^+_\e)^{1/2}}(0,\zeta_2)\rangle
=-\frac{2\kappa_0\kappa_2}{\kappa^2}\,\left(I_1(t)+ \frac{1}{\pi}\int_\RE
ds\,I_2(t,s)\,\right)\,\zeta_1^*\cdot\zeta_2\,,
$$
where
$$
I_1(t):=\frac{1}{2\pi i}\,\lim_{a\uparrow\infty}
\int_{-a}^a d\lambda\ e^{-it\lambda}\,\lambda
\left(f(\lambda)-f(-\lambda)\right)\,,
$$
$$
I_2(t,s):=\frac{1}{2\pi i}\,\int_\RE d\lambda\ e^{-it\lambda}\,\lambda^2
\frac{f(\lambda)-f(-\lambda)}{s^2+\lambda^2}\,,
$$
$$
f(\lambda):=
\frac{1}{\lambda+i\lambda_\e}\,\frac{
\frac{\e^2\lambda_\e}{3c^3}\,\lambda+i\,
\frac{m}{\lambda_\e}\,\left({\omega_0^2}+
\frac{\lambda_\e^2}{2}\right)}
{\frac{2\e^2}{3c^3}\, \lambda^2+i\,\frac{m\omega_0^2}{\lambda_\e^2}\,
(\lambda+i\lambda_\e)}\equiv 
\frac{1}{\lambda+i\lambda_\e}\,\frac{p(\lambda)}{q(\lambda)}\,.
$$
Then, by residue theorem and Jordan's lemma, when $\pm t>0$, one obtains 
$$
I_1(t)=\mp e^{\mp
\lambda_\e t}i\lambda_\e\,\frac{p(-i\lambda_\e)}{q(-i\lambda_\e)}
\pm e^{\mp \gamma_\e t}
\sum_{k=1,2}\frac{e^{\mp(-1)^ki\omega_\e t}z_k}{
z_k+i\lambda_\e}\,\frac{p(z_k)}{q'(z_k)}\,,
$$
where $z_k:=(-1)^{k}\omega_\e-\gamma_\e$, 
and
\begin{align*}
I_2(t,s)=&-\,\frac{e^{\mp
\lambda_\e t}\lambda_\e^2 }{s^2-\lambda_\e^2}\,
\frac{p(-i\lambda_\e)}{q(-i\lambda_\e)}
+ e^{\mp t\gamma_\e}
\sum_{k=1,2}\frac{z_k}{s^2+z_k^2}\,
\frac{e^{\mp(-1)^ki\omega_\e t}z_k}{
z_k+i\lambda_\e}\,\frac{p(z_k)}{q'(z_k)}\\
&+\,e^{\mp
|s|t}\,\frac{|s|}{2}\,
\left(\,\frac{1}{|s|\mp\lambda_\e}\,\frac{p(\mp i|s|)}{q(\mp i|s|)}
+\frac{1}{|s|\pm\lambda_\e}\,\frac{p(\pm i|s|)}{q(\pm i|s|)}\,
\right)
\end{align*}
Since
$$
\frac{1}{\pi}\int_\RE ds\,\frac{z_k}{s^2+z_k^2}=(-1)^k\,,
$$
one has
\begin{align*}
&I_1(t)+ \frac{1}{\pi}\int_\RE
ds\,I_2(t,s)\\
=&\mp e^{\mp\lambda_\e t}i\lambda_\e\,\frac{p(-i\lambda_\e)}{q(-i\lambda_\e)}
\pm e^{\mp \gamma_\e t}\,
\frac{e^{-i\omega_\e t}2z_k}{z_k+i\lambda_\e}\,
\frac{p(z_k)}{q'(z_k)}\\
&+\frac{1}{\pi}\int_\RE ds
\left(\,\frac{e^{\mp|s|t}|s|}{2(|s|-\lambda_\e)}\,
\frac{p(-i|s|)}{q(-i|s|)}
-\frac{e^{\mp\lambda_\e t}\lambda^2_\e}{s^2-\lambda_\e^2}\,
\frac{p(-i\lambda_\e)}{q(-i\lambda_\e)}\,\right)\\
&+\frac{1}{\pi}\int_\RE ds\,
\frac{e^{\mp|s|t}|s|}{2(|s|+\lambda_\e)}\,
\frac{p(i|s|)}{q(i|s|)}\,,
\end{align*}
where $k=2$ if $t>0$ and $k=1$ if $t<0$.
\end{proof}
\par \noindent
We distinguish three contributions in the previous formula for $I(t)$.
The one on the first line is due to the resonances of the system. In particular, 
we have a
purely exponential term with lifetime $\lambda_\e$, coming from the
projection on the stable subspace of the system, which dies out very
quickly; and terms exponentially damped, due to the complex
resonances. These last resonances are typical of the Breit-Wigner distribution,
and they are the only terms which survive to the usual single pole
approximation for the resolvent. They give both the breadth of
emission lines in the spectrum (or equivalently the lifetime of the
process) and the Lamb shift, as recalled in the
introduction. The other contributions to the survival amplitude
 need more direct analysis. To simplify the exposition we confine to the case
of $t>0$, the case $t<0$ being completely analogous.
The term $J_1(t)$ is nothing else that the Laplace
transform (which we denote by the symbol $\mathcal L$) a rational 
function. It is a tedious but standard calculation to verify that such a
Laplace transform is in fact the sum of a pure exponential and a
damped oscillation with the same characteristic exponents as the ones
coming from the first group of terms. So, the term $J_1(t)$
corresponds to a further resonance contribution.
\par\noindent
Concerning $J_2(t)$ the following result holds true.
\begin{lemma} There exist complex constants $c_1, c_2, c_3$ such that, for $t>0$ 
$$J_2(t)=\frac{1}{\pi} \mathcal L
\left(\,\frac{c_1s+c_2}{(s-\gamma_\e)^2 +
  \omega_e^2)}\,\right)+\frac{c_3}{\pi}P.V.\int^{\infty}_0 \frac{e^{-st}}{s-\lambda_\e}ds \,
$$
\end{lemma}
\begin{proof}
Introducing a parameter $\epsilon$ to isolate the
singularity in $\lambda_\e$, one can write
\begin{align*}
&J_2(t)=\frac{2}{\pi}\int_0^{+\infty} ds
\left(\,\frac{e^{-|s|t}|s|}{2(|s|-\lambda_\e)}\,
\frac{p(-i|s|)}{q(-i|s|)} -\frac{e^{-\lambda_\e
t}\lambda^2_\e}{s^2-\lambda_\e^2}\,
\frac{p(-i\lambda_\e)}{q(-i\lambda_\e)}\,\right)\\
=&\lim_{\epsilon\downarrow 0}\left(\, 
\frac{2}{\pi}\left(\int_0^{\lambda_\e - \epsilon}ds
\frac{e^{-|s|t}|s|}{2(|s|-\lambda_\e)}\,
\frac{p(-i|s|)}{q(-i|s|)} + \int_{\lambda_\e +\epsilon}^{+\infty} ds
\frac{e^{-|s|t}|s|}{2(|s|-\lambda_\e)}\,
\frac{p(-i|s|)}{q(-i|s|)}\,\right)\right.\\  
&\left.
-\frac{2}{\pi}\left(\int_0^{\lambda_\e - \epsilon} ds\frac{e^{-\lambda_\e
t}\lambda^2_\e}{s^2-\lambda_\e^2}\, 
\frac{p(-i\lambda_\e)}{q(-i\lambda_\e)}\,+ 
\int_{\lambda_\e +\epsilon}^{+\infty}\frac{e^{-\lambda_\e
t}\lambda^2_\e}{s^2-\lambda_\e^2}\,
\frac{p(-i\lambda_\e)}{q(-i\lambda_\e)}\,ds \right) \right)\,, 
\end{align*}
and the integrals in the second braces is vanishing for every $\epsilon$.
So, one remains with 
$$
J_2(t)=\frac{2}{\pi} P.V.\int_0^{\infty}
\frac{e^{-st}s}{2(s-\lambda_\e)}\, \frac{p(-is)}{q(-is)}\,
$$
Finally, the rational function in the above integral can be decomposed in a
singular (not integrable) part and an integrable one, obtaining 
$$
J_2(t)=\frac{1}{\pi} \mathcal L
\left(\,\frac{c_1s+c_2}{(s-\gamma_\e)^2 +  \omega_e^2)}\,\right)+
\frac{c_3}{\pi}P.V.\int^{\infty}_0 \frac{e^{-st}}{s-\lambda_\e}ds \,
$$
where $c_1$, $c_2$ and $c_3$ are constants depending on the parameters.
\end{proof}
The first integral has the time behaviour of the other already studied
contributions, and the last one is related to the exponential integral
function:
$$
\frac{c_3}{\pi}P.V.\int^{\infty}_0 \frac{e^{-st}}{s-\lambda_\e}ds =
-\frac{c_3}{\pi}e^{{-\lambda_\e}t} \Ei(\lambda_\e t)
$$
\par \noindent
From the well known asymptotic behaviour for large arguments of the
exponential integral function, one deduce that the leading
contribution for large times to the survival amplitude is of the order
$1\over t$ when $t\gg 1$. In summary, collecting the previous calculations,
we end with the following result 
\begin{theorem}
  There exist complex constants $c_1, c_2, c_3$ depending on the physical
  parameters $\e$, $m$, $c$, $\omega_0$ such that the survival amplitude 
is given by
$$
\mathcal S (t)=c_1 e^{-\lambda_\e |t|} +c_2 e^{-\gamma_\e
  |t|}e^{-i\omega_\e t} + c_3 e^{-\lambda_\e |t|} \Ei(\lambda_\e |t|)\,.
$$
\end{theorem}
\subsection{Transition amplitudes.}
Now we turn to the evaluation of the transition amplitudes.  As we saw
in the previous theorem, the unperturbed excited states of the
oscillator decay. The transition to a lower energy state take place
with the emission of electromagnetic radiation, photons in the Fock
space. To give a quantitative estimate of the probability of emission
of photons one has to decide which functions describe
the one particle states of the electromagnetic field one is interested
in.  We choose to
calculate the transition amplitudes in 
the case of a regularized plane wave of
(approximatively) given momentum.\par 
Classically, a natural choice for
the functions $\varphi$ representing the photons of energy $\hbar\nu$ 
should be the divergence-free plane waves of the kind
$$\varphi(x):=k \wedge \zeta\, e^{-i\nu k\cdot x/c}\,,\quad
|k|=1\,.
$$
Of course such functions are not square integrable and so
we consider the divergence-free regularization defined by
$$
\varphi^\epsilon(x):=
\left(k-\frac{i\epsilon}{\nu}\,\frac{x}{|x|}\right)
\wedge \zeta\, e^{-i\alpha k\cdot x/c-\epsilon|x|/c}\,,\quad
\nu>0\,,\, \epsilon>0\,,\, |k|=1\, .
$$
We choose this one as the 1-particle wavefunctions describing
photons. Eventually, we are interested in removing the cutoff
$\epsilon$ from the amplitude transitions.  
\par
We anticipate the definitions of some
quantities appearing in the statement and proof of the following
result. To evaluate the resolvent between states relevant to the
transitions, one need the scalar product
$\langle\varphi^\epsilon,{\mathcal G}^\pm_z\rangle$. Assuming 
that $k$ is along the
$z$-axis and $\zeta$ is along the $y$-axis, by  
elementary calculations one obtains
\begin{align*}
&\chi^\epsilon(\pm z):=\langle\varphi^\epsilon,{\mathcal
G}^\pm_z\rangle\\
&\equiv{k\wedge\zeta^*}\ \frac{1}{\alpha}\,
\left(\,\frac{\alpha^2-\epsilon(\epsilon\mp iz)}
{\alpha^2+(\epsilon\mp iz)^2}-\frac{i\epsilon}{2\alpha^2}\,
\ln\frac{\alpha+i\epsilon\pm z}{\alpha-i\epsilon\mp z}\,\right)\\
&\equiv
({k\wedge\zeta^*})(\chi^\epsilon_r (\pm z)+\chi^\epsilon_l (\pm z))\,.
\end{align*}
\par\noindent
We have written the function $\chi^\epsilon (\pm z)$ just defined, as the sum
$\chi_r^\epsilon (\pm z)+\chi_l^\epsilon (\pm z)$ with two summands, 
a rational one and
a logarithmic one. The first, $\chi^\epsilon_r (\pm z)$ has poles at the points
$\pm \alpha \pm i\epsilon$, and at the same points the logaritmic part
$\chi_l^\epsilon (\pm z)$ has branching points. Due to this fact, some cautions
are needed in the use of the residue theorem to evaluate transition
amplitudes. Note, moreover, that these branching points of $\chi$ are
to be thought as artificial byproducts of the regularization of the
plane wave, and that the poles of the function $\chi_r$ correspond to
the frequencies of the plane wave.  These poles depend on the
particular wavefunction representing the photon, at variance with the
poles of the function $1/q(z)$, which are determined by the physical
parameters of the system.  We indicate with $\mathcal C$ the
logarithmic cut in the complex plane, and we distinguish the
components of the cut with real part of a fixed sign ($\pm$), with
$\mathcal C^{\pm}$.  
With these premises, we state the following
\begin{lemma} 
For any $\zeta_1,\,\zeta_2\in \CO^3$
one has
$$
\langle(\varphi_1^\epsilon,0),e^{-it(L^+_\e)^{1/2}}(0,\zeta_2)\rangle
=\mathcal A^{\epsilon}(t)\,k\cdot(\zeta_1^*\wedge\zeta_2)\,,
$$
where
$$
\mathcal A^\epsilon(t)=-\frac{4\pi\e}{c  }\,\frac{\kappa_2}{\kappa^2}
\,I^\epsilon(t)
$$
and, with $\pm t>0$, 
\begin{align*}
&I^\epsilon(t)
=\mp e^{\mp
\lambda_\e t}\,\lambda_\e^2\,\frac{\chi(-i\lambda_\e)}
{q(-i\lambda_\e)}
\pm e^{\mp \gamma_\e t}
\frac{e^{-i\omega_\e t}2z_\pm}{z_\pm+i\lambda_\e}\,
\frac{2z_\pm+i\lambda_\e}{q'(z_\pm)}\,\chi(z_\pm)\\
&\pm e^{\mp \epsilon t}\,\frac{e^{\mp i\nu t}2(\pm\nu-i\epsilon)}
{\pm\nu+i(\lambda_\e-\epsilon)}\,
\frac{2\nu+i(\lambda_\e-2\epsilon)}{q(\pm\nu-i\epsilon)}\,
\frac{\mp\nu-i\epsilon}{2\nu}\\
&+\frac{i}{\pi}\int_\RE ds\,
\left(\,\frac{ e^{\mp
\lambda_\e t}\,\lambda_\e^3}{s^2-\lambda_\e^2}\,
\frac{\chi(-i\lambda_\e)}{q(-i\lambda_\e)}+
\,\frac{e^{\mp|s|t}|s|(-2|s|+\lambda_\e)}{2(|s|-\lambda_\e)}\,
\frac{\chi(-i|s|)}{q(-i|s|)}
,\right)\\
&+\frac{i}{\pi}\int_\RE ds\,
\frac{e^{\mp|s|t}|s|(2|s|+\lambda_\e)}{2(|s|+\lambda_\e)}\,
\frac{\chi(i|s|)}{q(i|s|)}\\
&+\frac{\epsilon}{\nu^3}\int_{\mathcal C} dz\, e^{-itz}\frac{2z+i\lambda_\e}
{z+i\lambda_\e}\,\frac{1}{q(z)}\\
&+\pi\,\frac{\epsilon}{\nu^3} \left(\int_{\mathcal C^{+}}dz\, e^{-itz} \frac{2z+i\lambda_\e}
{z+i\lambda_\e}\,\frac{z}{q(z)} - 
\int_{\mathcal C^{-}}dz\, e^{-itz} \frac{2z+i\lambda_\e}{z+i\lambda_\e}\,
\frac{z}{q(z)}\right)\,,
\end{align*} 
\end{lemma}
\begin{proof} One has, proceeding as in the proof of Lemma 3.1
\begin{align*}
&\langle(\varphi_1^\epsilon,0),(L_\e-z^2)^{-1}(0,\zeta_2)^+\rangle\\
=&\frac{4\pi c\,\e\kappa_2}{\kappa^2(\mp iz+\lambda_\e)}\,
\frac{(\pm 2iz-\lambda_\e)(\pm iz+\lambda_\e)}
{\left(\frac{2\e^2}{3c^3}\, z^2-\frac{m\omega_0^2}{\lambda_\e^2}\,
(\mp iz+\lambda_\e)\right)(\pm iz+\lambda_\e)}
\,\chi^\epsilon (\pm z)\,{(k\wedge\zeta_1^*)}\cdot\zeta_2\,.
\end{align*}
Therefore 
$$
\langle(\varphi_1^\epsilon,0),e^{-it(L^+_\e)^{1/2}}(0,\zeta_2)\rangle
=-\frac{4\pi\e}{c  }\,\frac{\kappa_2}{\kappa^2}\,\left(I_1(t)+ \frac{1}{\pi}\int_\RE
ds\,I_2(t,s)\,\right)\,(k\wedge\zeta_1^*)\cdot\zeta_2\,,
$$
where (we omit the dependence on $\epsilon$)
$$
I_1(t):=\frac{1}{2\pi i}\,\lim_{a\uparrow\infty}
\int_{-a}^a d\lambda\ e^{-it\lambda}\,\lambda
\left(g(\lambda)-g(-\lambda)\right)\,,
$$
$$
I_2(t,s):=\frac{1}{2\pi i}\,\int_\RE d\lambda\ e^{-it\lambda}\,\lambda^2
\frac{g(\lambda)-g(-\lambda)}{s^2+\lambda^2}\,,
$$
$$
g(\lambda):=
\frac{2\lambda+i\lambda_\e}{
\lambda+i\lambda_\e}\,\frac{\chi(\lambda)}
{q(\lambda)}\,.
$$
Now, let us choose a path in the complex lower halfplane wich has
the real axis as the upper side, avoids the cuts of the function
$\chi_l$ (say, the straight half lines parametrized as $z=\pm \nu +
u -i\epsilon\ , \pm u>0 $), and close itself at $\infty$ along a great
circle, as in the previous lemma. There will be contributions due to
the residues of the function $1/q(z)$, the residues the function
$\chi_r(z)$, and to the discontinuity of the logaritmic part of
$\chi_l(z)$ along the cut.  In the end, one obtains, when $\pm t>0$,
\begin{align*}
I_1(t)=&\mp e^{\mp
\lambda_\e t}\,\lambda_\e^2\,\frac{\chi(-i\lambda_\e)}
{q(-i\lambda_\e)}
\pm e^{\mp \gamma_\e t}
\sum_{k=1,2}\frac{e^{\mp(-1)^ki\omega_\e t}z_k}{z_k+i\lambda_\e}\,
\frac{2z_k+i\lambda_\e}{q'(z_k)}\,\chi(z_k)\\
&
\pm e^{\mp \epsilon t}\sum_{k=1,2}\,\frac{e^{\mp i\nu t}w_k}
{w_k+i\lambda_\e}\,
\frac{2w_k+i\lambda_\e}{q(w_k)}\,r_k + \int_{\mathcal C} \frac{2z+i\lambda_\e}{
z+i\lambda_\e}\,\frac{e^{-itz}}{q(z)}\,,
\end{align*}
where $z_\pm$ are the poles of $q$, 
 $w_k:=(-1)^{k}\nu-i\epsilon$
are the two poles of $\chi_r(z)$ and $r_k$ the residues of $\chi_r$, 
and finally $\mathcal C$ is the path along the cut of the logarithmic term.
\par\noindent
Moreover we have
\begin{align*}
&I_2(t,s)=i\,\frac{ e^{\mp
\lambda_\e t}\,\lambda_\e^3}{s^2-\lambda_\e^2}\,
\frac{\chi(-i\lambda_\e)}{q(-i\lambda_\e)}\\
&+e^{\mp \gamma_\e t}
\sum_{k=1,2}\frac{z_k}{s^2+z_k^2}\,
\frac{e^{\mp(-1)^ki\omega_\e t}z_k}{z_k+i\lambda_\e}\,
\frac{2z_k+i\lambda_\e}{q'(z_k)}\,\chi(z_k)\\
&
+e^{\mp \epsilon t}\sum_{k=1,2}\frac{w_k}{s^2+w_k^2}
\,\frac{e^{\mp i\nu t}w_k}
{w_k+i\lambda_\e}\,
\frac{2w_k+i\lambda_\e}{q(w_k)}\,r_k\\
&+i\,e^{\mp
|s|t}\,\frac{|s|}{2}\,
\left(\,\frac{\mp 2|s|+\lambda_\e}{|s|\mp\lambda_\e}
\,\frac{\chi(\mp i|s|)}{q(\mp i|s|)}
+\frac{\pm
2|s|+\lambda_\e}{|s|\pm\lambda_\e}\,\frac{\chi(\pm i|s|)}
{q(\pm i|s|)} \right)\,.\\
&+ \int_{\mathcal C}\, e^{-itz} \frac{2z+i\lambda_\e}
{z+i\lambda_\e}\,\frac{z^2}{q(z)}\frac{1}{s^2+z^2} \ .
\end{align*}
Therefore
one obtains
\begin{align*}
&I_1(t)+\frac{1}{\pi}\int_\RE ds\,I_2(t,s)\\
&=\mp e^{\mp\lambda_\e t}\,\lambda_\e^2\,\frac{\chi(-i\lambda_\e)}
{q(-i\lambda_\e)}
\pm e^{\mp \gamma_\e t}
\frac{e^{-i\omega_\e t}2z_k}{z_k+i\lambda_\e}\,
\frac{2z_k+i\lambda_\e}{q'(z_k)}\,\chi(z_k)\\
&\pm e^{\mp \epsilon t}\,\frac{e^{\mp i\nu t}2w_k}
{w_k+i\lambda_\e}\,
\frac{2w_k+i\lambda_\e}{q(w_k)}\,r_k\\
&+\frac{i}{\pi}\int_\RE ds\,
\left(\,\frac{ e^{\mp
\lambda_\e t}\,\lambda_\e^3}{s^2-\lambda_\e^2}\,
\frac{\chi(-i\lambda_\e)}{q(-i\lambda_\e)}+
\,\frac{e^{\mp|s|t}|s|(-2|s|+\lambda_\e)}{2(|s|-\lambda_\e)}\,
\frac{\chi(-i|s|)}{q(-i|s|)}
\,\right)
\end{align*}
\begin{align*}
&+\frac{i}{\pi}\int_\RE ds\,
\frac{e^{\mp|s|t}|s|(2|s|+\lambda_\e)}{2(|s|+\lambda_\e)}\,
\frac{\chi(i|s|)}{q(i|s|)}\\
&+\frac{\epsilon}{\nu^3} \int_{\mathcal C} dz\, e^{-itz}\frac{2z+i\lambda_\e}
{z+i\lambda_\e}\,\frac{1}{q(z)}\\
&+\pi \frac{\epsilon}{\nu^3} \left(\int_{\mathcal C^{+}}dz\, e^{-itz} \frac{2z+i\lambda_\e}
{z+i\lambda_\e}\,\frac{z}{q(z)} - 
\int_{\mathcal C^{-}}dz\, e^{-itz} \frac{2z+i\lambda_\e}{z+i\lambda_\e}\,
\frac{z}{q(z)}\right)
\end{align*}
where $k=2$ if $t>0$ and $k=1$ if $t<0$ and $\mathcal C^{\pm}$ are the
components of $\mathcal C$ with $\pm \text{\rm Sign(Re($z$))}>0$ respectively.

\end{proof}

Now we can give the time behaviour of the various terms, as in the
previous theorem.  A first group of terms is composed by the resonant
contributions. One distinguish natural resonances, depending
on the structural parameters only(the physical constants
$\e$, $m$, $c$), and the resonance due to the incident photon. For
$\epsilon \downarrow 0$ this last contribution reduces to a strictly
oscillating term, as expected.  A second group of terms is given by
the $s$ integrals. We give their behaviour for vanishing $\epsilon$ only.
The calculation is similar to the one already given for the survival amplitude, and one
has in the end Laplace transforms of rational functions with poles at
the resonances, producing other exponentials of the type already seen,
and an exponential integral function. So the leading behaviour of the
type $t^{-1}$ survives to the $\epsilon \downarrow 0$ limit.
\par\noindent The last
group of terms are the contour complex integrals.  These could be analized
asymptotically as Fourier integrals of rational functions, but they vanish
as $\epsilon\downarrow 0$. Summarizing we can state the following
\begin{theorem} There exist complex constants
$C_1$, $C_2$, $C_3$ depending on the physical parameters $m$,
$\e$, $c$, $\omega_0$, $\nu$, and a function $R(t)$ such that 
$$
\lim_{\epsilon\downarrow 0} \mathcal A^{\epsilon}(t)=C_1 e^{-\lambda_\e |t|} + 
C_2 e^{-\gamma_\e |t|}e^{-i\omega_\e t} + C_3 e^{-i\nu t}  + R(t)
$$
with $$R(t)=O({1}/{t})\,,\quad|t|\gg 1\,.$$ 
\end{theorem}
\appendix
\section{Quantization of abstract wave equations}

In this appendix we give a brief and self contained introduction to
the quantization of second order abstract wave equations, along the
lines traced by I.E.Segal in the fifties and sixties and with an emphasis
on the aspects of direct concern with our work. We refer for details
and different approaches to \cite{[Weiss]} and \cite{[Kato]} as
regards abstract wave equations and to \cite{[Segalrel]},
\cite{[Baez]} as regards quantization.  \par Let $B:D(B)\subseteq\H\to
\H$ be a injective self-adjoint operator on the real Hilbert space
$(\H, \langle\cdot,\cdot\rangle)$.  We denote by $\H^1$, the Hilbert
space given by the domain of $B$ with the scalar product
$\langle\cdot,\cdot\rangle_1$ leading to the graph norm, i.e.
$$
\langle\phi_1,\phi_2\rangle_1:=
 \langle B\phi_1,B\phi_2\rangle+\langle \phi_1,\phi_2\rangle\,.
$$
{\bf }\par
We then define the Hilbert space $\HB^1$ by completing 
the pre-Hilbert space $D(B)$ endowed with the the scalar product
$$
[\phi_1,\phi_2]_1:=
\langle B\phi_1,B\phi_2\rangle\,.
$$
We extend the self-adjoint operator $B$ to $\HB^1$ by considering
$\B:\HB^1\to\H$, the closed bounded extension of the densely
defined linear operator
$$
B:\H^1\subseteq\HB^1\to\H\,.
$$ 
Since $B$ is self-adjoint one has Ran$(B)^\perp$=Ker$(B)$, 
so that, being $B$ injective, Ran$(B)$ is dense in $\H$. Therefore we
can define $\B^{-1}:\H\to\HB^1$ as the closed bounded extension
of the densely defined linear operator
$$
B^{-1}:\text{\rm Ran$(B)$}\subseteq\H\to\HB^1\,.
$$
One can then verify that $\B$ is boundedly invertible with inverse
given by $\B^{-1}$. \par
Given $\B$ we can now introduce the space $\HB^2$ 
defined by
$$
\HB^2:=\left\{\phi\in\HB_1\ :\ \B\phi\in\H^{1}\right\}\,.
$$
On the Hilbert space $\HB^1\oplus\H$ with scalar product
$$
\langle\langle(\phi_1,\dphi_1),(\phi_2,\dphi_2)\rangle\rangle:=
\langle\B\phi_1,\B\phi_2\rangle+\langle\dphi_1,\dphi_2\rangle\,.
$$
we define the linear operator
$$
W_B:\HB_2\oplus\H^1\subseteq\HB^2\oplus\H\to \HB^1\oplus\H\,,\qquad 
W_B(\phi,\dphi):=(\dphi,-B\B\phi)\,.
$$
\begin{theorem} The linear operator $W_B$ is skew-adjoint.
\end{theorem}
Being $W_B$ skew-adjoint, by the Stone theorem it generates the strongly
continuous one-parameter group of isometric operators $e^{tW_B}$. By
defining $(\phi(t),\dphi(t)):=e^{tW_B}(\phi,\dphi)$, one has
that, in the case $\phi\in\HB^2$, $\dphi\in\H^1$,
$\phi(t)$ is the unique strong solution of the Cauchy problem
\begin{align*}&\ddot\phi(t)=-B\B\phi(t)\\
&\phi(0)=\phi\,,\quad
\dot\phi(0)=\dphi\,.
\end{align*}
Given an arbitrary real Hilbert space $(\Ko,
\langle\cdot,\cdot\rangle_\Ko)$ we will denote by $(\Ko_c,
\langle\cdot,\cdot\rangle_c)$ its standard complexification, 
i.e. $\Ko_c=\Ko\oplus\Ko$, the
multiplication by the complex unity being defined by $i\psi:=J\psi$, 
$J(\varphi_1,\varphi_2):=(-\varphi_2,\varphi_1)$, and
$$
\langle\psi_1,\psi_2\rangle_c:=
\langle\psi_1,\psi_2\rangle_{\K\oplus\K}
-i\langle\psi_1,J\psi_2\rangle_{\K\oplus\K}\,.
$$
Given any linear operator $A:D(A)\subseteq\Ko\to\Ko$ on $\Ko$, we define
$A_c:D(A_c)\subset \Ko_c\to\Ko_c$ by $D(A_c):=D(A)\times D(A)$,
$A_c(\varphi_1,\varphi_2):=(A\varphi_1,A\varphi_2)$. Conversely, given
any linear operator $L:D(L)\subset\Ko_c\to\Ko_c$,
$L(\varphi_1,\varphi_2)\equiv
(L_1(\varphi_1,\varphi_2),L_2(\varphi_1,\varphi_2))$, 
we define
$L_r:D(L_r)\subseteq\Ko\to\Ko$ by $D(L_r):=
\left\{\phi\in\K\,:\,(\phi,0)\in D(L)\right\}$, $L_r:=L_1(\varphi,0)$.\par
By the above definitions $B_c$ is self-adjoint and $(W_B)_c=W_{B_c}$, where  
$$
W_{B_c}:\HB^2_c\oplus\H_c^1\subseteq\HB_c^2\oplus\H_c\to \HB_c^1\oplus\H_c\,,\qquad 
W_{B_c}(\psi,\dot\psi):=(\dot\psi,-B_c\B_c\psi)\,.
$$
By the Stone formula one has the following
\begin{lemma}
\begin{align*}
&e^{tW_B}\\
=&\lim_{a\uparrow\infty}\lim_{\epsilon\downarrow 0}\,\frac{1}{2\pi i}
\int_{-a}^a d\lambda\ e^{-it\lambda}((iW_{B_c}-(\lambda+i\epsilon))^{-1}-
(iW_{B_c}-(\lambda-i\epsilon))^{-1})_r\,.
\end{align*}
\end{lemma}
\noindent
The following lemma translates the wave flow in a Schr\"odinger like one.
\begin{lemma} The map 
$$
C_B:\HB^1\oplus\H\to\H_c\,,\qquad C_B(\phi,\dphi):=(\B\phi,\dphi)\,.
$$
is unitary once one makes $\HB^1\oplus\H$ a complex Hilbert space
by introducing the complex structure 
$J_B(\phi,\dphi):=(-\B^{-1}\dphi,\B\phi)$ and by defining
$i(\phi,\dphi):=J_B(\phi,\dphi)$ and the scalar product 
$$
\langle\langle(\phi_1,\dphi_1),(\phi_2,\dphi_2)\rangle\rangle_B:=
\langle\langle(\phi_1,\dphi_1),(\phi_2,\dphi_2)\rangle\rangle-i
\langle\langle(\phi_1,\dphi_1),J_B(\phi_2,\dphi_2)\rangle\rangle\,.
$$ 
Moreover
$$
e^{tW_B}=C_B^*e^{-itB_c}\,C_B\,.
$$ 
\end{lemma}
\noindent
Once we have transformed the abstract
wave equation $\ddot\phi(t)=-B\B\phi(t)$, defined on the real Hilbert
space $\HB^1\oplus\H$, into the Schr\"odinger equation
$i\dot\psi=B_c\psi$, defined on the complex Hilbert space $\H_c$, we can
then (second) quantize it in the standard way.
Let us define $\K$ to be the bosonic Fock space over
$\H_c$ (see \cite{[RSI]}, section II.4, for the definition). 
For any $\psi\in\H$ we define the self-adjoint operator on
$\K$ given by the Segal field 
$\S(\psi):=\frac{1}{\sqrt 2}(\C(\psi)+\C^*(\psi))$, were $\C$ and $\C^*$
denote the usual creation and destruction operators (see
\cite{[RSII]}, 
section X.7, for the definition). Given the Segal
field $\S$ we can then define the Weyl system $\W(\psi):=e^{i\S}\psi$, so that
$$
\W(\psi_1+\psi_2)=e^{\frac{i}{2}\,\langle\psi_1,\psi_2\rangle_c}
\W(\psi_1)\W(\psi_2)\,.
$$ 
The unitary strongly continuous one-parameter group of evolution on
$\K$ defined by $\U(t):=\Gamma(e^{-itB_c})$ satisfies the relations
$$
\W(e^{-itB_c}\psi)=\U(t)\W(\psi)\U(t)^*
$$
and we denote by $d\Gamma(B_c)$ the self-adjoint operator on $\K$ with
generates $\U(t)$ (we refer to 
\cite{[RSII]}, section II.4, and \cite{[Cook]}
for the definitions of $\Gamma$ and $d\Gamma$). 
The quantum hamiltonian corresponding to 
the (second) quantization of the abstract
wave equation $\ddot\phi(t)=-B\B\phi(t)$ is defined by
$\Ham:=\hbar\, d\Gamma(B_c)$.
Suppose now that we start with a selfadjoint operator $A$ on the real
Hilbert space $\Ko$. Denoting by $A^+$ the positive part of $A$ and by
$\Ko^+$ the projection of $\Ko$ onto the positive spectral subspace. 
Then we can apply the previous construction to $B:=(A^+)^{1/2}$,
considered as an
injective self-adjoint operator on $\H:=\Ko^+$. However, since 
$$
\langle\psi_1,e^{-it(A^+)^{1/2}_c}\psi_2\rangle=
\langle\psi_1^+,e^{-it(A^+)^{1/2}_c}\psi_2^+\rangle\,,
$$  
where $\psi^+$ denotes the projection of $\phi$ on $\Ko^+$, we 
will work with $d\Gamma((A^+)^{1/2}_c)$ on the bosonic Fock space over $\Ko$.

\end{document}